# Carbon Schwarzites Behavior Under Ballistic Impacts

Levi C. Felix[1,2], Cristiano F. Woellner[3], Douglas S. Galvao[1,2]

[1] 'Gleb Wataghin' Institute of Physics, State University of Campinas, Campinas-SP, 13083-970, Brazil

[2] Center for Computational Engineering & Sciences, State University of Campinas, Campinas-SP, 13083-970, Brazil

[3] Physics Department, Federal University of Paraná, Curitiba-PR, 81531-980, Brazil

ABSTRACT

Schwarzites are 3D crystalline porous materials exhibiting the shape of Triply Periodic Minimal Surfaces (TPMS). They possess negative Gaussian curvature, created by the presence of rings with more than six sp2-hybridized carbon atoms. Recently, new routes to their synthesis have been proposed. Due to its foam-like structure, schwarzites are interesting for mechanical energy absorption applications. In this work, we investigate through fully atomistic reactive molecular dynamics the mechanical response under ballistic impacts of four structures from primitive (P) and gyroid (G) families (two structures within each family). The two structures in the same family differ mainly by the ratio of hexagons to octagons, where this ratio increases the 'flatness' of the structures. Although the penetration depth values are higher in the 'flatter' structures (P8bal and G8bal), the absorbed kinetic energy by them is considerably higher, which yields them a better energy-absorption performance.

INTRODUCTION:

Schwarzites (Figure 1 and Table 1) are 3D crystalline porous materials exhibiting the shape of Triply Periodic Minimal Surfaces (TPMS). They possess negative Gaussian curvature, created by the presence of rings with more than six sp2-hybridized carbon atoms and the 'flatness' of these curvatures is dependent on the ratio of the hexagons to non-hexagon rings [1]. Recently, new routes to their synthesis have been proposed [2,3]. Due to its foam-like structure, schwarzites are interesting for mechanical energy-absorption applications [4].

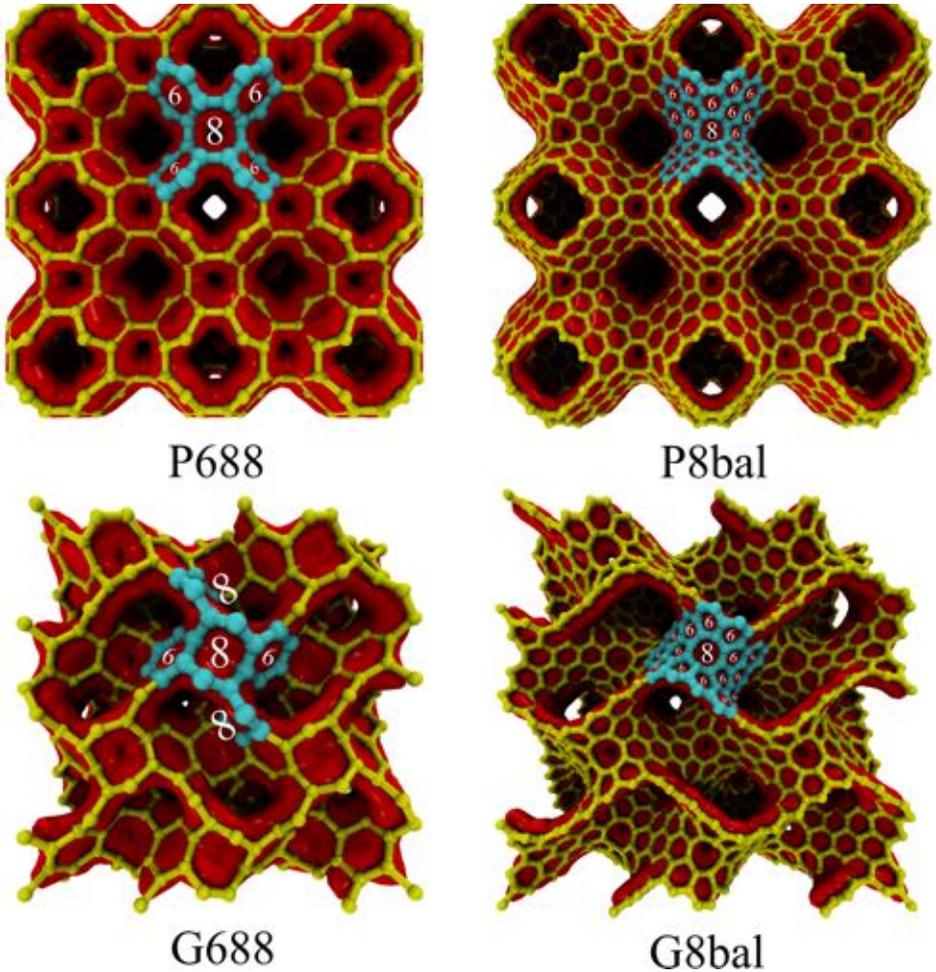

**Figure 1**: Schwarzite structures investigated in this work.

**METHODOLOGY:**

In this work, we have investigated the mechanical response of four structures from primitive (P) and gyroid (G) families (two structures within each family) under ballistic impacts through Molecular Dynamics (MD) simulations. We have carried out fully atomistic MD simulations using the Adaptive Intermolecular Reactive Empirical Bond-Order (AIREBO) force field [5], as implemented in the open-source code LAMMPS [6]. The systems were initially thermalized using an NVT ensemble for 100 ps at 100 K. The ballistic impact simulations (Figure 2) were performed using an NVE ensemble, where the projectile was simulated by a Lennard-Jones potential ($\varepsilon_0 = 1.0$ eV and $\sigma_0 = 7.5$ Å) with the same mass as the target structure. Finite boundary conditions and a time step of 0.1 fs were used in all simulations. The VMD software [7] was used to visualize MD trajectories and snapshots.

Table 1: Structural properties of the four schwarzite structures: # of atoms per cubic cell (n), # of atoms in the supercell (N), lattice parameter (a), specific mass (ρ) and cubic size (L).

| Structure | n | N | a [Å] | ρ [g/cm³] | L [nm] |
|---|---|---|---|---|---|
| P688 | 48 | 3072 | 7.8 | 2.0 | 3.31 |
| P8bal | 192 | 5184 | 14.9 | 1.2 | 4.51 |
| G688 | 96 | 2592 | 9.6 | 2.2 | 2.89 |
| G8bal | 384 | 3072 | 18.4 | 1.2 | 3.72 |

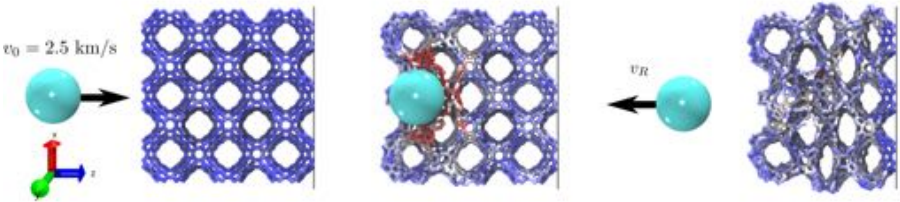

Figure 2: Representative MD snapshots of a ballistic impact simulation of a Lennard-Jones particle against a P8bal structure (cross-section view). An initial projectile velocity ($v_0$) of 2.5 km/s was set in all simulations. The projectile bounced back in all cases, where $v_R$ is the returning velocity of the projectile after the impact.

**RESULTS AND DISCUSSION:**

From Figure 3, we obtain the maximum penetration depth values (d) and the returning velocity ($v_R$) acquired by the projectile after impact. The kinetic energy (KE) absorbed by the structure ($KE_{abs}$) is the difference between the initial and returning KE of the projectile, which in turn should be equal to the work (W) on the projectile by the impact process [8] (given by the area under the curves shown in Figure 4 for each structure). The values of d, $v_R$, $KE_{abs}$ and W are shown in Table 2. In Figure 5,6,7 and 8 we present representative MD snapshots for the impact simulation on a P688, P8bal, G688 and G8bal structures, respectively.

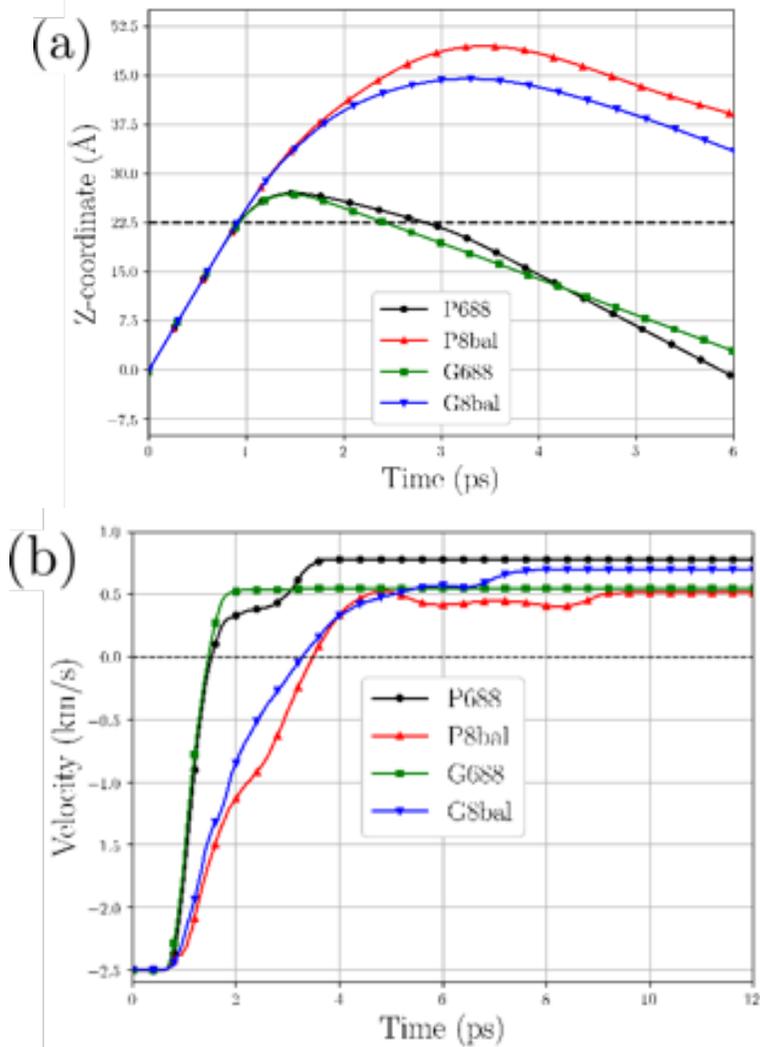

Figure 3: (a) Z-coordinate of the projectile as a function of time for all four structures, where the maximum of the curves represent the turning points at the maximum penetration depth in the structures and the dashed line indicates where the projectile hits the structure. (b) The velocity of the projectile as a function of time, where all curves reach a constant value, which represents the returning velocity value of the projectile on each structure.

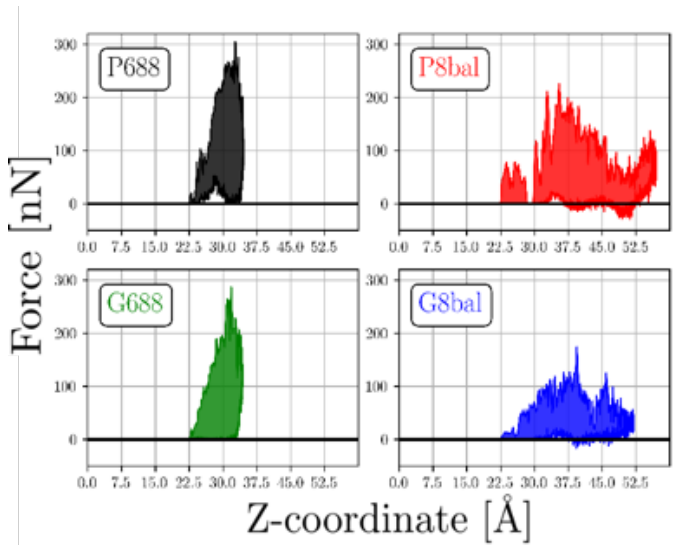

Figure 4: Force on the projectile as a function of its coordinate. The area of these curves gives the work realized by the structures on the projectile during the impact process.

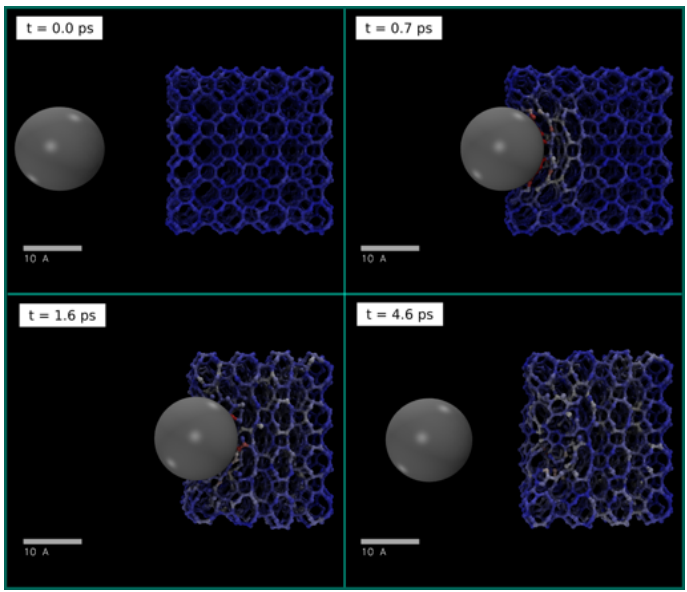

Figure 5: Representative MD snapshots of the trajectory of the impact simulation for the P688 structure. The colors on atoms represent local stress values, from low (blue) to high (red) values.

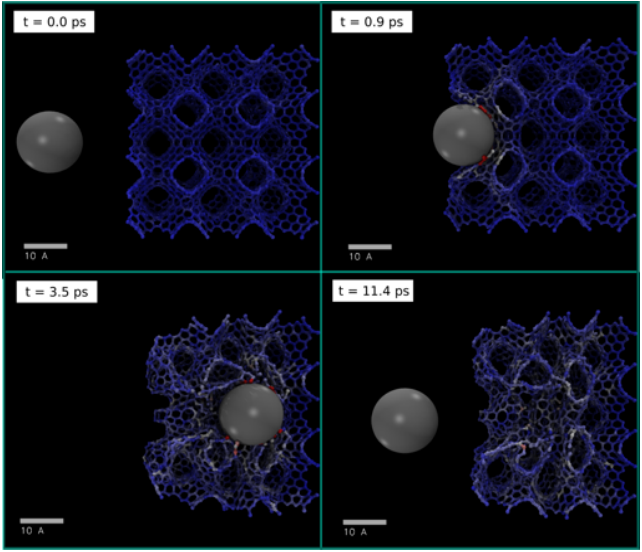

Figure 6: Representative MD snapshots of the trajectory of the impact simulation for the P8bal structure. The colors on atoms represent local stress values, from low (blue) to high (red) values.

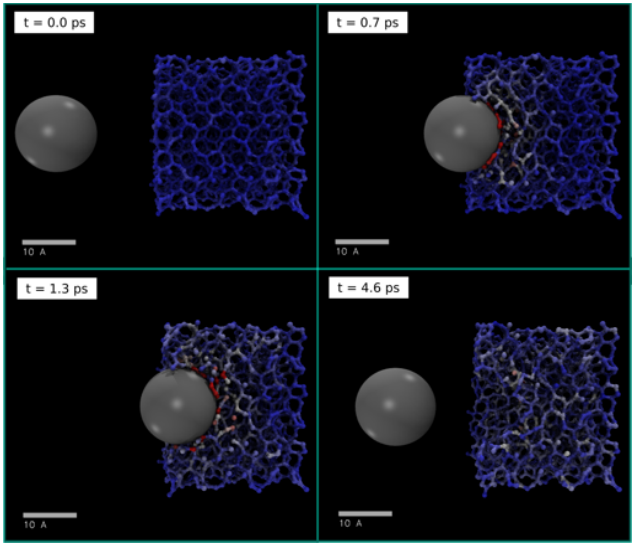

Figure 7: Representative MD snapshots of the trajectory of the impact simulation for the G688 structure. The colors on atoms represent local stress values, from low (blue) to high (red) values.

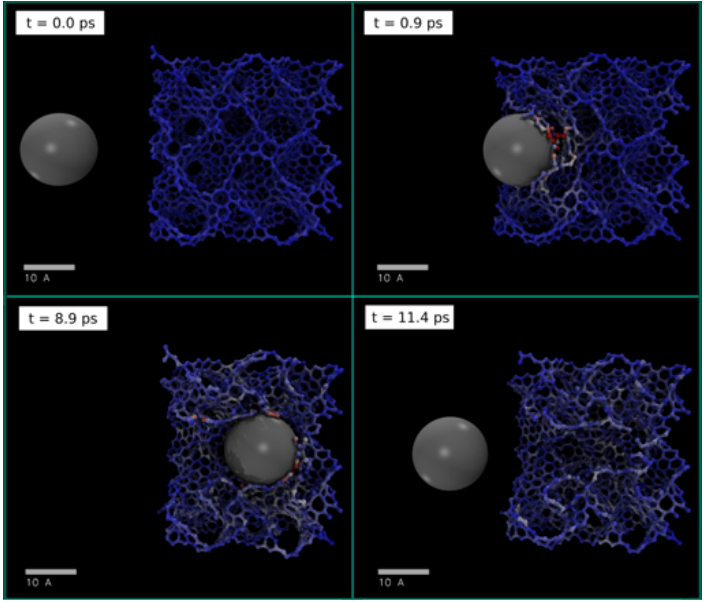

**Figure 8: Representative MD snapshots of the trajectory of the impact simulation for the G8bal structure. The colors on atoms represent local stress values, from low (blue) to high (red) values.**

Table 2: Energy absorption performance of all four schwarzites structures: Penetration depth (d), returning velocity ($v_R$), absorbed kinetic energy ($KE_{abs}$), work done on the projectile (W) and the percentage error between $KE_{abs}$ and W.

| Structure | d [Å] | $v_R$ [km/s] | $KE_{abs}$ [aJ] | W [aJ] | Error between $KE_{abs}$ and W [%] |
|---|---|---|---|---|---|
| P688 | 4.5 | 0.8 | 172.61 | 172.64 | 0.02 |
| P8bal | 26.9 | 0,5 | 308.44 | 310.44 | 0.65 |
| G688 | 4.3 | 0.6 | 153.55 | 154.66 | 0.72 |
| G8bal | 22.0 | 0.7 | 176.29 | 175.84 | 0.26 |

## CONCLUSIONS:

We have investigated the trends of kinetic-impact (ballistic) performance of two families of schwarzites (Primitive and Gyroid) through fully atomistic reactive molecular dynamics (MD) simulations. Our MD results showed that schwarzites are very effective to absorb kinetic energy but their performance decreases as their 'flatness' (dependent on the ratio of the hexagons to non-hexagon rings) decreases. Therefore, the 'flatter' structures (P8bal and G8bal) have a better energy-absorption performance under kinetic impact. For P8bal and G8bal the projectile penetration is about four times deeper than in P688 and G688, with more extensive structural fractures. Considering that it was already demonstrated that macro models of these materials can be 3D printed [9], our results open new perspectives to create new or to improve schwarzite-based functional engineered materials [10].